\documentclass[10pt,preprint]{article}          
\usepackage{amsmath}    
\usepackage{graphicx}   
\usepackage{endnotes}   

\title{A toolkit to describe and interactively display three-manifolds embedded in four-space}

\author{Don V. Black\footnote{IEEE Computer Society - Orange County,CA; email:dblack@ieee.org}}

\date{ March 10, 2010}

\begin{document}
\maketitle
\begin{abstract}

A data structure and toolkit are presented here that allow for the description and manipulation of mathematical models of three-manifolds and their interactive display from multiple viewpoints via the OpenGL 3D graphics package. The data structure and vector math package can be extended to support an arbitrary number of Euclidean spatial dimensions.

A model in 4-space is described by its bounding pure simplicial 3-complex.  By intersecting a 3-flat with this 3-manifold, the algorithm will extract the requested closed pure simplicial 2-complex surface enclosing the desired 3D slice. The user can interactively rotate, pan, zoom, and shade arbitrary 3D solid or wire-frame views of the revealed 3D object created by intersection, thus exploring both expected and unexpected symmetries or asymmetries in the world of 3-manifolds in 4-space.

\end{abstract}

\setcounter{tocdepth}{6}

\section{Introduction}
\label{sec:intro}

In order to understand complex, multi-variate simulations of phenomena, it has proven to be useful to be able to observe the shape, symmetries and asymmetries of higher dimensional mathematical models which emerge from these studies. This capability allows the observer insights into the nature of complex phenomena such as: viewing relativistic spacetime interactions; preventing collision in the design of a robotic arm or fine tuning a robotic assembly line; exploring intersecting brane models in physics; discovering unexpected elegant symmetries in physical laws.

Describing and viewing extra-dimensional objects on the computer has been a thorny problem due to a lack of an elegant extensible data structure to represent higher dimensional models as well as the processing requirements of an interactive solution.

Techniques for viewing four- and higher-dimensional objects have been proposed and in some cases implemented by Noll\cite{Noll:1965} as early as 1965, and later beginning in the 1990's by
Hollasch\cite{Hollasch:1991},
Hanson\cite{Hanson:1991,Hanson:1994b,Chu:2009},
Banks\cite{Banks:1992},
and in the domain of visualizing four-dimensional spacetime by 
Weiskopf\cite{Weiskopf:2001,Weiskopf:2005b}.

Presented here are a fundamental data structure and attendant library of tools to represent and display a 4D model as a four-manifold with boundary by describing a mesh of its 3D bounding simplices. The toolkit as described and demonstrated here can both generate the bounding three-simplices\footnote{\textbf{$m$-simplex:} a convex hull of (m+1) independent points in an $n$-dimensional ($n$D) Euclidean space $E^{n}$ (where $n\geq m$). A triangle is a two-simplex, while a tetrahedron is a three-simplex. A simplex need not be regular.} bordering the mathematical 4D model at a specified level-of-detail, and allow the user to interactively explore this model in real-time by selecting 3D projections and intersections while viewing the resultant 3D object just as does an architect, engineer, or  video gamer.

\section{Implementation}

While the vector library was implemented in seven Euclidean dimensions, only four dimensions are described and demonstrated here. Each 4D object is represented by its bounding three-manifold which is a closed surface\footnote{\textbf{Closed surface} - "A surface is closed if it is compact, connected and has no boundary; in other words it is a compact, connected Hausdorff space in which each point has a neighborhood homeomorphic to the plane."\cite{Armstrong:1983}. This definition will be used, with suitable extension, to describe the hyper-surface.} approximated by a pure simplicial 3-complex\footnote{\textbf{pure simplicial $m$-complex} - a set of aligned non-intersecting simplices of dimension $m$ in $n$-space, where $m \leq n$.}
in 4-space.

\begin{figure}[ht]
  \centering
  \includegraphics[width=0.75\linewidth]{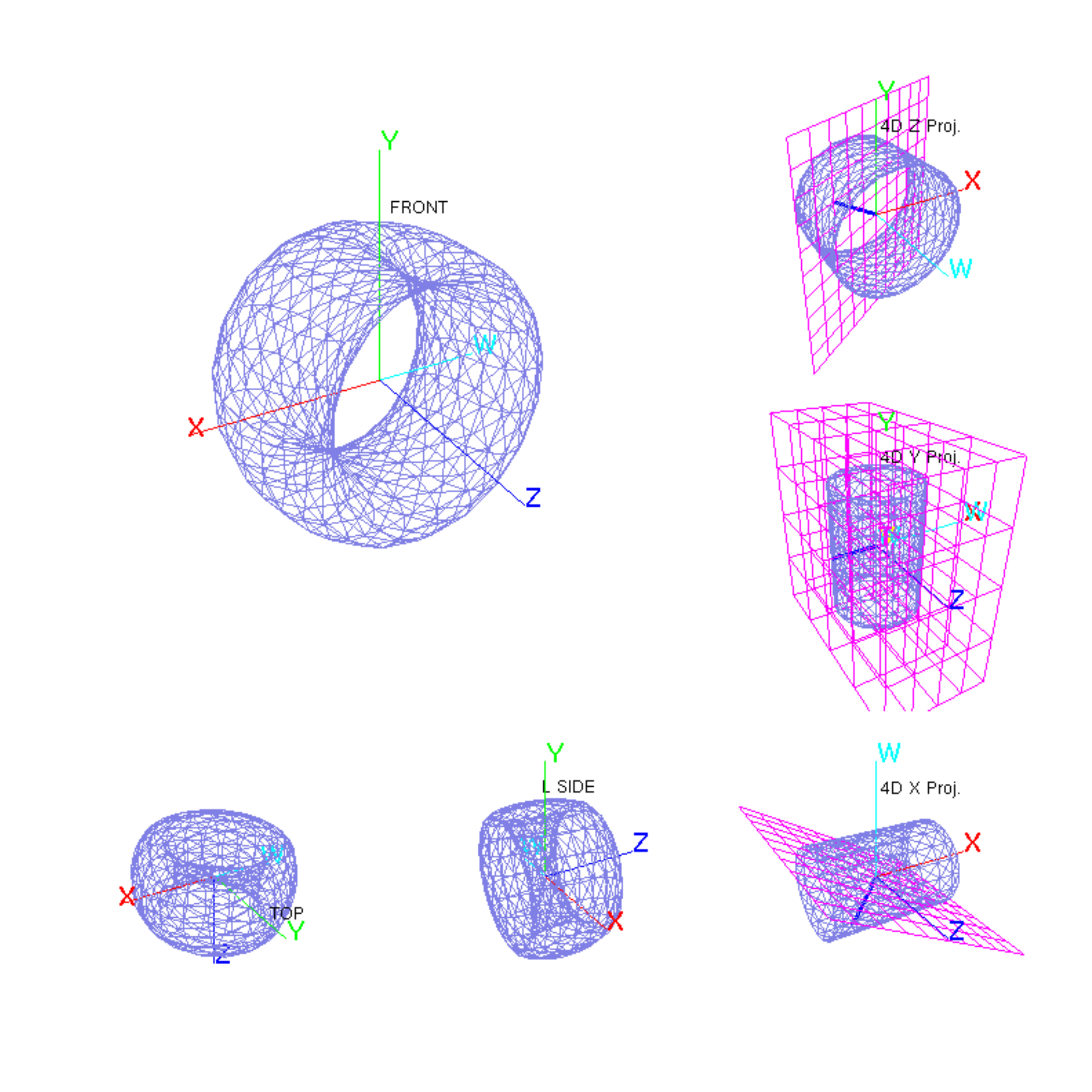}
  \caption{\label{fig:2TorusProj}  Two-torus in 4D and projected obliquely onto a three-flat}
  \scriptsize{(Note that a projection onto the 3-flat is shown, not an intersection with the 3-flat)}
\end{figure}

This paper describes a technology to reveal the embedded 3D structures composed of 2-manifolds within a 3-manifold in 4-space generated by intersecting a 3-flat with the model's pure simplicial 3-complex, and displaying the resultant closed 2-manifold as a conventional bounded 3D object in 3-space via the OpenGL library.

The vertex database and vector math module were implemented in 7D to support a future application that will require 7-space. The impact on performance of raising the matrices from 4x4 to 7x7 was deemed to be acceptable for a proof-of-concept. This decision facilitated the inclusion of a sample projection from 5D of a 4-flat intersecting a manifold with extent in 5 dimensions as described in Section~\ref{sec:resu} and depicted in Figure~\ref{fig:5DObject}.

A 4-manifold with boundary model is created and described by its bounding closed 3-manifold in the same manner as a bounded 3-manifold is described by its bounding 2-manifold. A sphere enclosed by its surface of bounding triangles is an example of the latter.

The bounding closed 3-manifold is a pure simplicial 3-complex of adjacent tetrahedra sharing faces, edges, and vertices.  The data structure for each tetrahedron is a list of four vertices. Since all vertices are shared among the 4 adjacent tetrahedra in the 3-complex, the vertices are merged into one indexed vertex array.  The tetrahedron data structure element contains a list of the four indices into this vertex array.

A 3-flat\footnote{\textbf{3-flat}: an isotropic unbounded 3D hyperplane in Euclidean $n$-space $E^n$, where $n\ge 3$} hyperplane equation is manipulated interactively via a 3D icon in a 4D GUI. The resulting 3-flat of infinite extent, as depicted by the shaded rectilinear grid in the right column of Figure~\ref{fig:2TorusProj}, is then intersected with the edges of each tetrahedron,
thus extracting a triangle (or a pair of coplanar triangles) which can be displayed as a surface element of the closed 2-manifold that defines the extracted 3D object. An animated video of the gridded icon moving through the depicted 4D object and the resulting 3D projection can be found online.\cite{Black:2009d}

This extracted 3D object can then be passed on to the OpenGL library and the usual flat or smooth shading effects implemented.

\begin{figure}[p]           
\begin{quote}               
\begin{center}
\textbf{\emph{Pseudo-Code of Sample Data Structures}}
\end{center}
\vspace{-16.0pt}
\begin{tabbing}             
$~~~~~~$\=$~~~~~~$\=$~~~~~~$\=$~~~~~~$\=$~~~~~~$\=$~~~~~~$\=$~~~~~~$\=$~~~~~~~~~$\=$~~~~~~~~~~~~~~~~~$ \\
$struct~Vec7~\{     $\>\>\>\>\>\>\>\>$  //~7D~double~precision~vector$   \\
    \>$double~$\>\>\>$t,x,y,z,w,v,u;$\>\>\>\>$//~7~components~of~vector$ \\
$\};$                                                                    \\
\\
$struct~Object~\{$\>\>\>\>\>\>\>\>$//~Linked~list~and~OpenGL~color~info$    \\
    \>$OBJ\_TYPE $\>\>\>$ objType;          $\>\>\>\>$//~Triangle~or~Tetrehedra$                 \\
    \>$Material $\>\>\>$  material;         $\>\>\>\>$//~OpenGL~material~color/lighting~info$  \\
    \>$Object $\>\>\>$    *next;            $\>\>\>\>$//~Link~to~next~singly~linked$-$list~object$ \\
$\};$                                                                             \\
\\
$struct~Triangle:public~Object~\{$\>\>\>\>\>\>\>\>$//~2D~Triangle~Object~of~3~vertices$     \\
$~~~~vecArray~*Verts;$                                                               \\
$~~~~int~iA,~iB,~iC;$\>\>\>\>\>\>\>\>$//~Indices~of~3~vertices~into~*Verts~array$      \\
$\};$                                                                               \\
                                                                                        \\
$struct~Tetrahedron:public~Object~\{$\>\>\>\>\>\>\>\>$//~3D~Tetrahedron~Object~of~4 vertices$        \\
$~~~~vecArray~~~~*Verts;$   \\
$~~~~int~~~~~~~~~iVert[4];$\>\>\>\>\>\>\>\>$//~Indices~of~4~vertices~into~*Verts~array$\\
$\};$  \\
 \\
$struct~vecArray~\{$                \\
$~~~~int    $\>\>\>$vecLen;$\>\>\>\>\>$//~Number~of~used~entries$              \\
$~~~~int    $\>\>\>$maxLen;$\>\>\>\>\>$//~Allocated~array~size$              \\
$~~~~int    $\>\>\>$*fixed;$\>\>\>\>\>$//~flags~and~counts~for~sums~\&~average$              \\
$~~~~Vec7   $\>\>\>$*vecPtr;$\>\>\>\>\>$//~The~array~of~7D~vectors$              \\
$~~~~double $\>\>\>$fClose;$\>\>\>\>\>$//~Used~to~merge~shared~fClose~vertices$              \\
$\};$

\end{tabbing}
\end{quote}
\centering
\caption{\label{fig:CodeObject}Structures used to slice 4D tetrahedra into displayable 3D triangles.}
\scriptsize{The Tetrahedron is sliced into one or two Triangles. The Object structure contains the lighting model data required by OpenGL.  One global 7D vecArray is shared by all 4D objects and a second is shared by all the sliced 3D displayable objects.  Vertices of adjacent $n$D simplices are identical, so most vertices are shared.}
\end{figure}

\subsection{Data Structure}

As shown in the pseudo-code description of the data structure in Figure~\ref{fig:CodeObject}, the 3-manifold is a list of tetrahedra and their vertices.  Since the 4 vertices of adjacent 3-simplices are shared, only approximately one vertex per simplex need be stored in the 3-manifold's data set.  The number of simplices required to describe an $m$-manifold increases exponentially with the dimension $m$. Thus the data structure and computational complexity also increase exponentially with the manifold's dimensionality $m$.

\subsection{Object-Generator}
The Object-Generator projects the user-specified 4D model's bounding closed pure simplicial 3-complex onto the 3-sphere enclosing the origin of the world's 4-space at a user specified radius.  For example, a 4D hyper-torus can be displayed at an arbitrary scale.  A balanced binary search tree (BBST) algorithm merges proximate vertices to reduce storage and subsequent computational complexity.  A snippet of code for the generation of a three-torus is provided in Appendix~\ref{sec:objectgenerator} and is described as follows.

The make3torus() method iteratively computes the hyper-spherical coordinates of the projected position of the cube onto the three-sphere. The threeTorusXYZW() method converts to rectilinear components, and putTetras() tessellates the cube's eight vertices into six tetrahedra. The decomposition of a cube into six tetrahedra is described by Black.\cite{Black:2007}
The verticesArray.putVert() method in putTetra() builds the BBST lexically from the components of the vertices. The resulting vertices and tetrahedra list are output to an ASCII text file.

\begin{figure}[ht]
  \centering
  \includegraphics[width=1.0\linewidth]{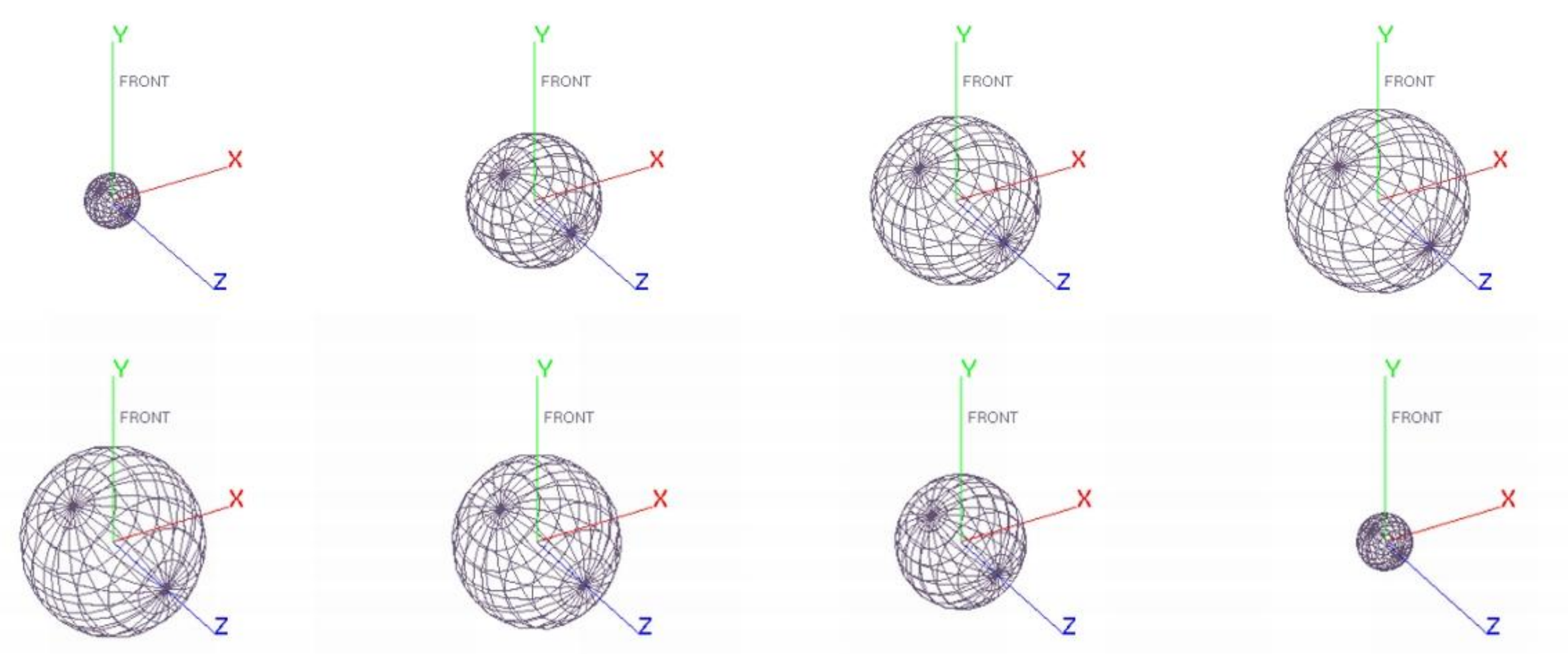}
  \caption{\label{fig:4SphereSliceW}A 4D Sphere sliced eight times along the W axis by a 3D hyperplane.}
  \scriptsize{The sliced $\mathcal{S}^2$'s diameter changes with the $\mathcal{S}^3$'s diameter in the classic Flatland\cite{Abbot:1884} manner.}
\end{figure}

\subsection{Object-Viewer}
The user can manipulate the grid-like 3D icon as shown in Figure~\ref{fig:2TorusProj} via a conventional "click-and-drag" GUI to position and orient the 3-flat in 4-space resulting in a user-defined 3-flat equation.  The GUI description is beyond the scope of this paper.

\paragraph{Intersection}
As noted above, the 3-flat via its equation is intersected with each tetrahedron.
The structure of the 3-manifold in 4D will be visualized as a sequence of 2D surfaces. So ideally we would intersect the 3-flat with this 3-manifold and get a 2D surface. By moving the 3-flat we can get different slices of  this 3 manifold. But since this 3-manifold we are considering, is a simplicial 3-complex, it consists of linear simplices of known face, edge, and vertex connections between these 3-simplices. Using this information, the intersection operation can be simplified to just performing the intersection of the edges of the 3-simplices with the 3-flat, and connecting these intersection points according to the connectivity of their parent simplices to get a 2D mesh. The 2D mesh is then rendered via OpenGL.

A discussion of the development of the plane equation in 4D can be found in Hanson\cite{Hanson:1994b} or Chu \& Hanson\cite{Chu:2009}. The intersection is simplified to a solution of the 3-flat hyperplane equation with the line equation of each edge of the tetrahedron.  The resulting 3 or 4 vertices define one or two triangles that are one face of the bounding 2-manifold of the new 3D object.  This new object is then rendered using OpenGL. A snippet of code for viewing the 3D bounds of a 4D object is provided in Appendix~\ref{sec:objectviewer} and is described as follows.

The list of vertices and enclosing tetrahedra with indices into the vertex list are input from the ASCII text file described above. A codimension 3 of 4 hyperplane equation
is generated from the user's manipulation of the mouse and keyboard to position and orient the 3-flat icon\footnote{The \textbf{3-flat icon} is shown in Figure~\ref{fig:2TorusProj} as a 2D or 3D rectilinear grid depending on its 4D to 3D projection.} within the multi-view projection of the 4D object's bounding manifold, as shown in Figure~\ref{fig:2TorusProj}.

Following the code in Appendix~\ref{sec:objectviewer} from bottom to top, clipScene() selects each object type data element in the scene linked-list and passes it to clipObject() where the object type is determined.\footnote{Many more object types were included during the research and development phase. For simplicity, the object type is limited to the Tetrahedron in this paper.} The clipNdxTet() method is used to clip the six edges of the Tetrahedron via isLineIn3Flat() and xsctLineTo3Flat().\footnote{The six intersections can be parallelized into six independent threads.} Since the 3-flat is a 3D hyperplane of infinite extent, ignoring degenerate cases\footnote{Degenerate Tetrahedron clipping cases are those which intersect only one vertex, are coincident with only one edge or one face, or are trivially-in.}, three or four coplanar vertices will be intersected by the 3-flat. Three vertices form a triangle, while four vertices can be tesselated into two coplanar adjacent triangles as shown by \emph{case 3:} and \emph{case 4:}, respectively, in the code snippet. These vertices, now internal to the clipping 3D hyperplane's 3-space, are thus 3D vertices describing a 3D object's boundary.

The resultant Triangles as described by their 3D vertices are inserted into a new Scene linked-list and handed off to the OpenGL library for interactive display.

\paragraph{Projection}
    A user-specified 4-space component is discarded and a 3D object is composed of the remaining 3 components. The 3D object is derotated to the 3-flat view screen, and the object's resultant 3D vertices are handed to OpenGL for 3D display.\cite{Hanson:1994b}

\begin{figure}[ht]
  \centering
  \includegraphics[width=1.0\linewidth]{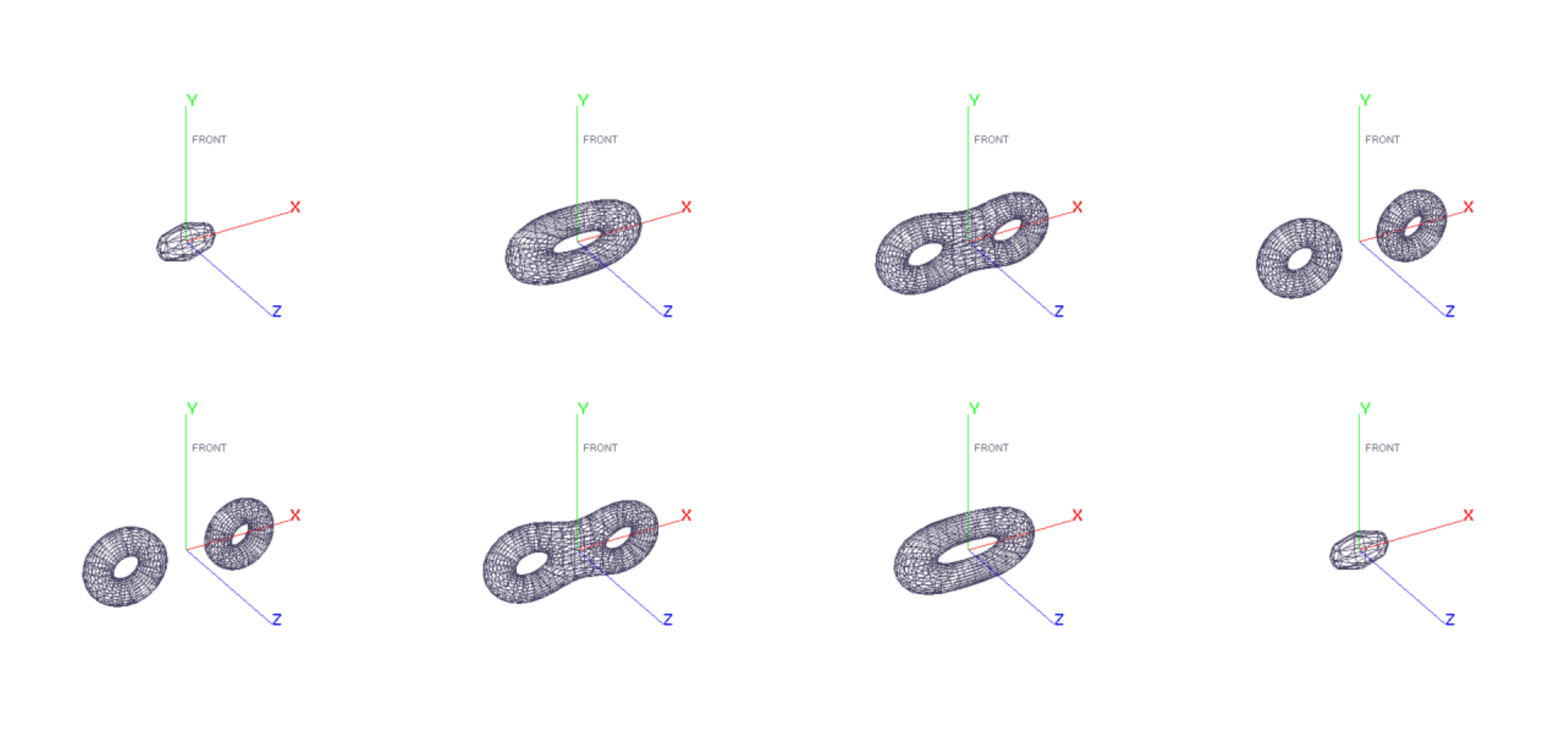}
  \caption{\label{fig:4TorusSliceW03}  A 4D Torus sliced eight times along the W-axis by a 3D hyperplane}
\end{figure}

\section{Results}
\label{sec:resu}

These examples have been selected to illustrate the capabilities of the described method.

Figure~\ref{fig:4SphereSliceW} shows eight frames from an animation of the 3-flat progressing along the $W$-axis of a 4D three-sphere. The radius of the resultant two-sphere increases up to the maximum radius of the three-sphere and then decreases. An animated video of this sequence can be found online\cite{Black:2009b}.

Figure~\ref{fig:4TorusSliceW03} shows eight frames from a more interesting animation of the 3-flat progressing along the $W$-axis of a three-torus which was defined to be $\mathcal{T}^3\rightarrow\mathcal{R}^4$ as mapped in Appendix~\ref{sec:objectgenerator}. In this case, dimensional reduction reveals hidden 3D structure and symmetry within the 3-torus.  As the 3-flat progresses along the $W$-axis, the revealed object transforms from genus zero, to genus one, to genus two and back. Some interesting symmetries can be observed in the animation sequences that can be found online\cite{Black:2009c}.

\begin{figure}[ht]
  \centering
  \includegraphics[width=1.0\linewidth]{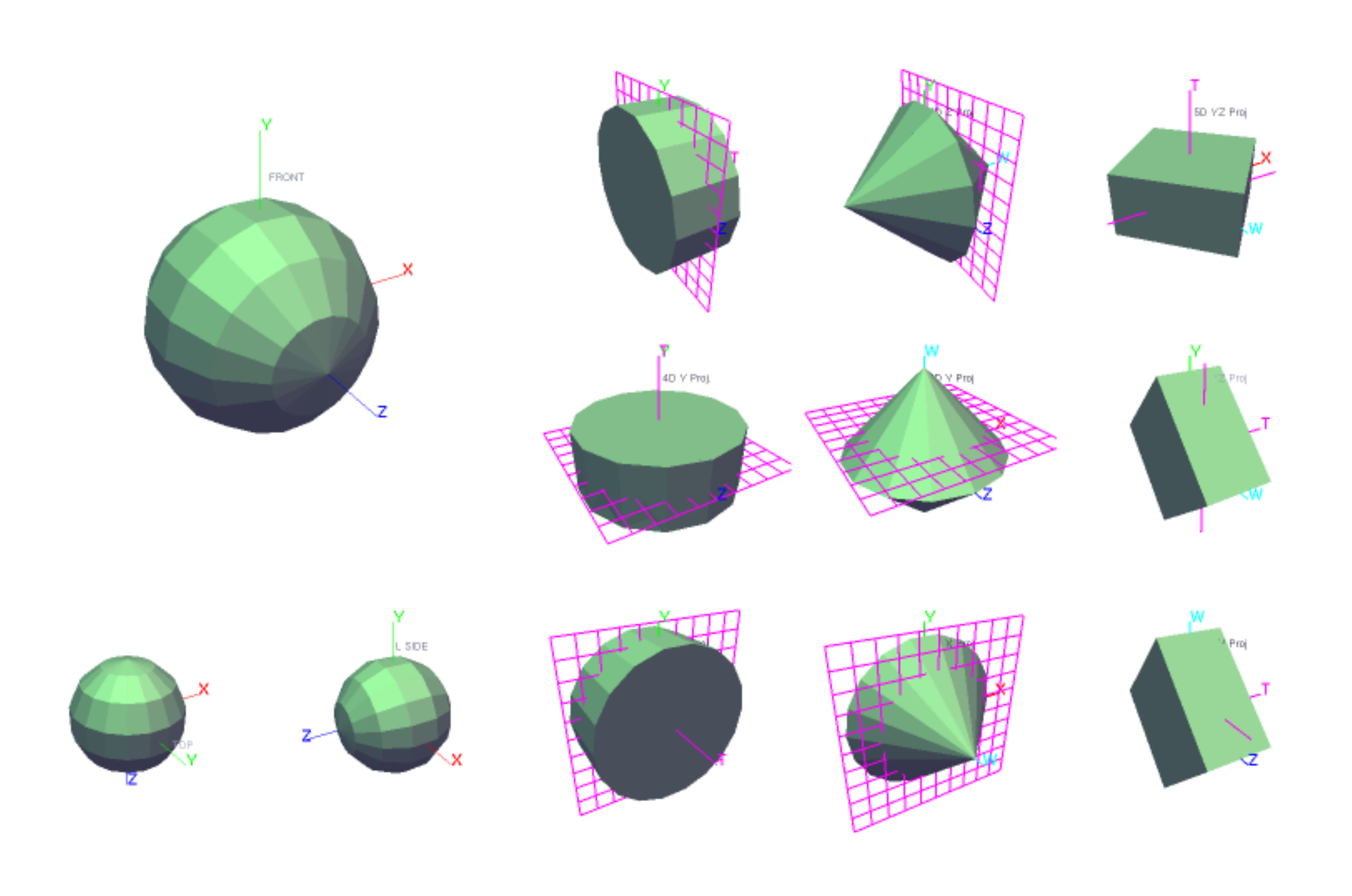}
  \caption{\label{fig:5DObject}  A 3-Manifold Spanning 5-Space Depicted in Twelve 3D Viewports}
  \scriptsize{Col~1: Three 3D views clockwise: (Front,Side,Top). The remaining views are projections along axis pairs into 3-space: Col~2:(WX,WY,WZ). Col~3:(TX,TY,TZ). Col~4: (ZY,ZX,YX). Note the 3D rectilinear grid icon is projected as a 1D line in Column~4.}
\end{figure}


As shown in Figure~\ref{fig:5DObject}, the algorithm can be extended to include higher dimensional objects.  Shown here is a 4D three-manifold homeomorphic to a three-sphere that has been extruded along the $t$-axis to yield an object spanning five dimensions.  As suggested by Figure~\ref{fig:5DObject}, the algorithm lends itself to the creation and exploration of an interactive Minkowski spacetime diagram in four dimensions or more spatial or temporal dimensions.

\section{Discussion}

A complex three-manifold in Euclidean four-space can be interactively explored in real-time on a desktop computer. While only two four-dimensional objects were demonstrated here, the algorithm and implementation support higher dimensional objects as is shown in Figure~\ref{fig:5DObject}. The data structure and algorithm are designed such that they can be extended to allow visualization of objects of yet greater spatial dimensionality.
The 5D example suggests that twice the number of 2D viewports are required for each additional dimension displayed, depending on the art of the developer. While exploring the 4D objects is interactive on the author's desktop\footnote{AMDx2 at 2.6GHz w/ 4GB RAM and OTS NVidia graphics hardware running XP-64.}, the interactivity of the 5D implementation can be described at best as near-realtime.  Since the nature of the algorithm will allow for a parallel implementation on contemporary multi-GPU devices, such implementation will likely address the 5D performance issue.

\vspace{24pt} \noindent\textbf{Acknowledgments.}\small \quad The author would like to thank the following individuals for their contributions used in this paper: Dr. Richard Palais for his three-torus definition; Dr James Arvo for his Toytracer Raytrace code; Digital ChoreoGraphics and the Edutech Foundation for their computer hardware and infrastructure.


\appendix

\newpage

\section{Appendix - Source Snippet: Object-Generator}

\label{sec:objectgenerator}
The Data Structures used in the following pseudo-code are defined in Figure~\ref{fig:CodeObject}.

\begin{verbatim}

//  make3torus -> threeTorusXYZW()
//  make3torus -> putTetras() -> putPrism() -> putTetra() -> printf()

//  External method to create a RedBlack Binary Search Tree from va, inserting v
extern int RedBlackBST( struct vecArray va[LARGE_NUMBER], Vec7 v);

struct vecArray
{
    int putVert( Vec7 v) { RedBlackBST( verticesArray[], v));
    Vec7 vert;             // 7 components: (t,x,y,z,w,v,u)
} // vecArray

struct vecArray verticesArray[LARGE_NUMBER];

//  putPrism's velocity vector is Vtx0->Vtx3, when Vvec is 0.

void putTetra( Vec7 Vtx0, Vec7 Vtx1, Vec7 Vtx2, Vec7 Vtx3, Vec7 Vvec, Vec7 Ovec)
{
    int ndxV = verticesArray.putVert( Vvec),    ndxO = verticesArray.putVert( Ovec);
    int ndx0 = verticesArray.putVert( Vtx0),    ndx1 = verticesArray.putVert( Vtx1),
        ndx2 = verticesArray.putVert( Vtx2),    ndx3 = verticesArray.putVert( Vtx3);
    printf("itetrahedra %d,%d,%d,%d,%d,%d\n",   ndx0, ndx1, ndx2 ,ndx3, ndxV, ndxO);
    putColor( l_color, l_color);
} // putTetra

//  Output three adjacent 3D Tetrahedra to form a Prism (extruded triangle).
//  Prism knows that V0->V3, V1->V4, V2->V5 are velocity vectors for the vertices
void putPrism( Vec7 Velocity, Vec7 V0, Vec7 V1, Vec7 V2, Vec7 V3, Vec7 V4, Vec7 V5)
{
    putTetra( V0, V1, V2, V3, Velocity, Vec7());   // Prism 1, Tet 1
    putTetra( V1, V2, V3, V4, Velocity, Vec7());   // Prism 1, Tet 2
    putTetra( V2, V3, V4, V5, Velocity, Vec7());   // Prism 1, Tet 3
} // putPrism

//  Output two adjacent 3D Prisms to form a Cube
void putTetras( Vec7 Velocity, Vec7 V0, Vec7 V1, Vec7 V2, Vec7 V3,
                               Vec7 V4, Vec7 V5, Vec7 V6, Vec7 V7)
{
    putPrism( Velocity, V0, V1, V2, V4, V5, V6);    // Prism 1
    putPrism( Velocity, V1, V3, V2, V5, V7, V6);    // Prism 2
} // putTetras
\end{verbatim}
\newpage
\begin{verbatim}

//  Compute the 5D components of the 7D vector
Vec7 threeTorusXYZW( double Phi, double Psi, double Theta,
                     double radius = 5.0, double tube = 2.0, double depth=1.0)
{
        return Vec7( 0.0,
                    (radius + (tube + depth * cos(Phi)) * cos(Psi)) * cos(Theta),
                    (radius + (tube + depth * cos(Phi)) * cos(Psi)) * sin(Theta),
                              (tube + depth * cos(Phi)) * sin(Psi),
                                      depth * sin(Phi)
                    );
} // threeTorusXYZW

//  Generate a 3-torus in 7-space. Rotate about a hypersphere center
make3torus(double deltaAng, double radius = 5.0, double tube
= 2.0, double depth=1.0) {
    double Phi0 = 0.0;
    for( Phi=deltaAng; Phi<=2*PI; Phi += deltaAng)
    {
        double Psi0 = 0.0;
        for( Psi=deltaAng; Psi <= 2*PI; Psi+=deltaAng)
        {
            double Theta0 = 0.0;
            for( Theta=deltaAng; Theta<=2*PI; Theta += deltaAng)
            {

                OldRow0 = threeTorusXYZW( Phi0, Psi0, Theta0, radius, tube, depth);
                OldRow1 = threeTorusXYZW( Phi0, Psi,  Theta0, radius, tube, depth);
                OldRow2 = threeTorusXYZW( Phi0, Psi0, Theta,  radius, tube, depth);
                OldRow3 = threeTorusXYZW( Phi0, Psi,  Theta,  radius, tube, depth);
                NewRow0 = threeTorusXYZW( Phi,  Psi0, Theta0, radius, tube, depth);
                NewRow1 = threeTorusXYZW( Phi,  Psi,  Theta0, radius, tube, depth);
                NewRow2 = threeTorusXYZW( Phi,  Psi0, Theta,  radius, tube, depth);
                NewRow3 = threeTorusXYZW( Phi,  Psi,  Theta,  radius, tube, depth);

                putTetras(Velocity, OldRow0, OldRow1, OldRow2, OldRow3,
                                        NewRow0, NewRow1, NewRow2, NewRow3);
                Theta0 = Theta;
            }
            Psi0 = Psi;
        }
        Phi0 = Phi;
    } // for
    //  List sorted vertices vertices:
    verticesArray.output();
} // make3torus

\end{verbatim}

\newpage

\section{Appendix - Source Snippet: Object-Viewer}

\label{sec:objectviewer}
The Data Structures used in the following pseudo-code are defined in Figure~\ref{fig:CodeObject}.
\begin{verbatim}
//  clipScene() -> clipObject() -> clipNdxTet() -> xsctLineTo3Flat() -> isLineIn3Flat()

extern vecArray slicedArray;        // Indexed list of 7D vertices

//! \brief isLineIn3Flat - cofactors were precomputed for this 3-flat.

int isLineIn3Flat(Slice3D &hyperBrane, Vec7 La, Vec7 Lb, Vec7 *rslt)
{
    Vec7 Col5n      = Vec7(1.0, La.t, La.x, La.y, La.z);
    double n        = hyperBrane.cofactors * Col5n;
    Vec7 vec        = Lb - La;
    Vec7 Col5d      = Vec7(0.0, vec.t, vec.x, vec.y, vec.z);
    double d        = hyperBrane.cofactors * Col5d;

    if( REAL_NEAR0(d))   {
        if( bF02 && !!REAL_NEAR0(n))        return ST_IN3FLAT_ALL;
        else                                return ST_IN3FLAT_NONE;
    }
    double s = - n/d ;
    *rslt = La + s*vec;
    return SMALL_RANGE( s)?ST_IN3FLAT_XSCT
                          :ST_IN3FLAT_NONE;
} // isLineIn3Flat

int xsctLineTo3Flat(Slice3D &hyperBrane, int iLine0, int iLine1)
{
    //      P1 should have been the 0th vertex: too many edges for 3D!

    Vec7    Q0 = Vec7();
    Vec7    Vtx0 = *((slicedArray.vecPtr)+iLine0);
    Vec7    Vtx1 = *((slicedArray.vecPtr)+iLine1);
    int     iQ0  = -1;

    int     in3Flat = isLineIn3FlatA(  hyperBrane, Vtx0, Vtx1, &Q0)
    switch(in3Flat) {
        case ST_IN3FLAT_XSCT:   // Add vertex to segment list
                                iQ0 = slicedArray.putVec( Q0);
                                return iQ0;
        case ST_IN3FLAT_ALL:    // Mark this segment as Trivially In
                                return 0;
        default:                // No intersection
        case ST_IN3FLAT_NONE:
                                break;
    }
    return -1;
} // xsctLineTo3Flat

Object * clipNdxTet( Slice3D &brane, iTetrahedra *iTet) {
    Object * obj = NULL,
           *next = NULL;
    int j, rc[6], iVert[6][2], iV[6];
    /* A */ iVert[0][0] = iTet->iVert[0];   iVert[0][1] = iTet->iVert[1];
    /* B */ iVert[1][0] = iTet->iVert[0];   iVert[1][1] = iTet->iVert[2];
    /* C */ iVert[2][0] = iTet->iVert[0];   iVert[2][1] = iTet->iVert[3];
    /* D */ iVert[3][0] = iTet->iVert[1];   iVert[3][1] = iTet->iVert[2];
    /* E */ iVert[4][0] = iTet->iVert[1];   iVert[4][1] = iTet->iVert[3];
    /* G */ iVert[5][0] = iTet->iVert[2];   iVert[5][1] = iTet->iVert[3];

            //  ******************************************  //
            //  Try to intersect all six edges
            //  ******************************************  //

    int icnt=0;
    for( j=0; j<6; j++)
    {
        rc[j] = xsctLineTo3FlatA_MathW( brane, iVert[j][0], iVert[j][1]);
        if(rc[j]>0) iV[icnt++] = rc[j];
    } // for

            //  ******************************************  //
            //  Check Trivial Out
            //  ******************************************  //

    if( (rc[0]<0) && (rc[1]<0) && (rc[2]<0) &&
        (rc[3]<0) && (rc[4]<0) && (rc[5]<0))    return NULL;

            //  ******************************************  //
            //  Is it a triangle?
            //  This is only a test to draw Triangles
            //  since the vertices are unordered!!!!
            //  This must be coordinated with the iTetrahedra
            //  vertex generation code (mkhyper.cpp).
            //  ******************************************  //

    switch(icnt) {                  // Number of edges: should be 3 or 4!
    case 3:     obj = new iTriangle( &slicedArray, iV[0], iV[1], iV[2]);
                if(bF03) obj->material.diffuse = SEG_PURP;
                break;
    case 4:     obj = new iTriangle( &slicedArray, iV[0], iV[1], iV[2]);
                if(bF03) obj->material.diffuse = SEG_BLU;
                obj->next = new iTriangle( &slicedArray, iV[1], iV[2], iV[3]);
                if(bF03) obj->next->material.diffuse = SEG_GRN;
                break;

    case 0:     obj = NULL;
                break;
    case 1:     obj = new iEdge( &slicedArray, iV[0], iV[0]);
                if(bF03) obj->material.diffuse = SEG_MINT;
                break;
    case 2:     obj = new iEdge( &slicedArray, iV[0], iV[1]);
                if(bF03) obj->material.diffuse = SEG_MINT;
                break;
    case 5:     obj = new iPolygon( &slicedArray, iV[0], iV[1], iV[2], iV[3], iV[4]);
                if(bF03) obj->material.diffuse = SEG_RED;
                // I don't think this can happen:
                break;
    }
    return obj;
} // clipNdxTet


Object * clipObject( Object *object)
{
    extern Object * clipNdxTet( Slice3D &brane, iTetrahedra *iTet);
    Vec7        hyperPoint  = Vec7(sliceRange.t.min,
                                   sliceRange.x.min, sliceRange.y.min, sliceRange.z.min,
                                   sliceRange.w.min, sliceRange.v.min, sliceRange.u.min);
    extern vecArray vectorArray, slicedArray;
    vecArray *ptrArray = &slicedArray;

          switch( object->objType)
          {
          default:
              printf("clipObject: Invalid objType (%d): '%s'\n",
                            object->objType, sName[object->objType]);
              break;

          case OBJ_TYPE_ITET:       // Tetra must go to (degenerate or 3,4 or 5 verts) face
              return clipNdxTet( hyperPlane, (struct iTetrahedra *) object);
              break;

          } // switch
          return NULL;
} // clipObject

\end{verbatim}
\newpage
\begin{verbatim}

Scene * clipScene( Scene &pScene)
{
      extern vecArray vectorArray;
      extern vecArray slicedArray;
      extern vecArray normalArray;
      Object        *obj = NULL;
      Scene         *newScene   = new Scene( &pScene, &slicedArray, &normalArray);
      Object        *newObject  = NULL,
                    *prvObject  = NULL,
                    *nextObj;

      int       oIndex = 0;

      time( &l_timeCopyBeg);
      iCopyScene++;         // Count passes thru copyScene
      copyArray( &slicedArray, &vectorArray);

      nextObj = pScene.first;
      while( nextObj && (NULL == (newObject = clipObject( nextObj))))
          nextObj = nextObj->next;

      //    an ITET could return many faces (up to 2)
      while(newObject->next)    newObject = newObject->next;

      //    Got at least one valid object, carry on
      oIndex++;
      for( obj=nextObj->next; obj != NULL; obj=obj->next)
      {
          oIndex++;     // printf("%3d %s\n", oIndex++, sName[obj->type]);
          prvObject = newObject;
          newObject = clipObject( obj);
          //    Skip this object if a NULL comes back
          if( newObject) {
              prvObject->next = newObject;
              //    ITET could return many faces
              while(newObject->next)    newObject = newObject->next;
          } else {
              newObject       = prvObject;
          }
      } // for
    return newScene;
} // clipScene

\end{verbatim}

\newpage
\bibliographystyle{unsrt}

\end{document}